\documentclass{article}
\usepackage{spconf,amsmath,graphicx}

\usepackage[utf8]{inputenc} 
\usepackage[T1]{fontenc}    
\usepackage{hyperref}       
\usepackage{url}            
\usepackage{booktabs}       
\usepackage{amsfonts}       
\usepackage{nicefrac}       
\usepackage{microtype}      

\usepackage{color}
\usepackage{graphicx}
\usepackage{amsmath}
\usepackage{dsfont}
\usepackage{amssymb}
\usepackage{amsfonts}
\usepackage{multirow}

\title{Visualizing Missing Surfaces In Colonoscopy Videos using Shared Latent Space Representations}

\name{Shawn Mathew$^{1 \star}$ \qquad Saad Nadeem$^{2 \star}$ \thanks{$\star$ Equal Contribution} \qquad Arie Kaufman$^{1}$}
\address{$^1$ Stony Brook University, Department of Computer Science, USA\\
     $^2$ Memorial Sloan Kettering Cancer Center, Department of Medical Physics, USA}
\begin{document}
%
\maketitle

\begin{abstract}
Optical colonoscopy (OC), the most prevalent colon cancer screening tool, has a high miss rate due to a number of factors, including the geometry of the colon (haustral fold and sharp bends occlusions), endoscopist inexperience or fatigue, endoscope field of view, etc. We present a framework to visualize the missed regions per-frame during the colonoscopy, and provides a workable clinical solution. Specifically, we make use of 3D reconstructed virtual colonoscopy (VC) data and the insight that VC and OC share the same underlying geometry but differ in color, texture and specular reflections, embedded in the OC domain. A lossy unpaired image-to-image translation model is introduced with enforced shared latent space for OC and VC. This shared latent space captures the geometric information while deferring the color, texture, and specular information creation to additional Gaussian noise input. This additional noise input can be utilized to generate one-to-many mappings from VC to OC and OC to OC. The code, data and trained models will be released via our Computational Endoscopy Platform at \url{https://github.com/nadeemlab/CEP}.
\end{abstract}


\section{Introduction}
\begin{figure}[t!]
\begin{center}
\setlength{\tabcolsep}{2pt}
\begin{tabular}{cccc}
\includegraphics[width=0.11\textwidth]{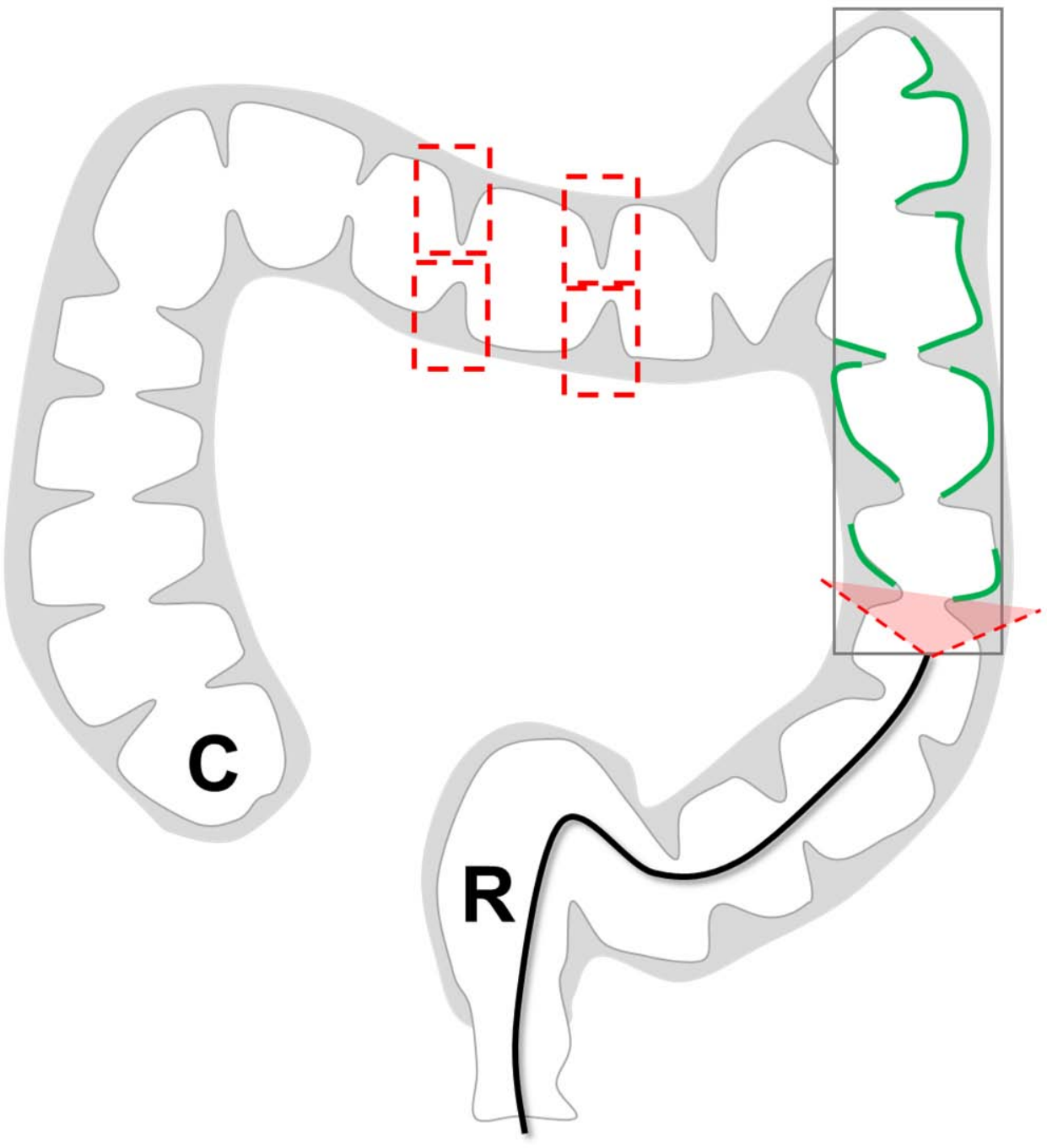}&
\includegraphics[width=0.11\textwidth]{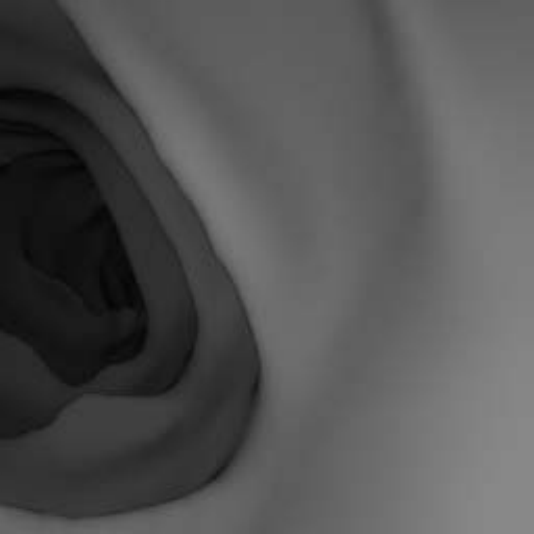}&
\includegraphics[width=0.11\textwidth]{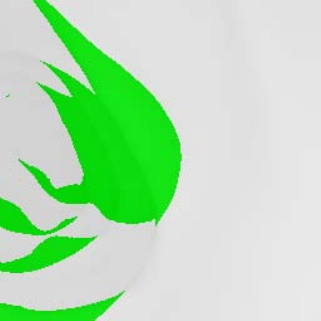}&
\includegraphics[width=0.11\textwidth]{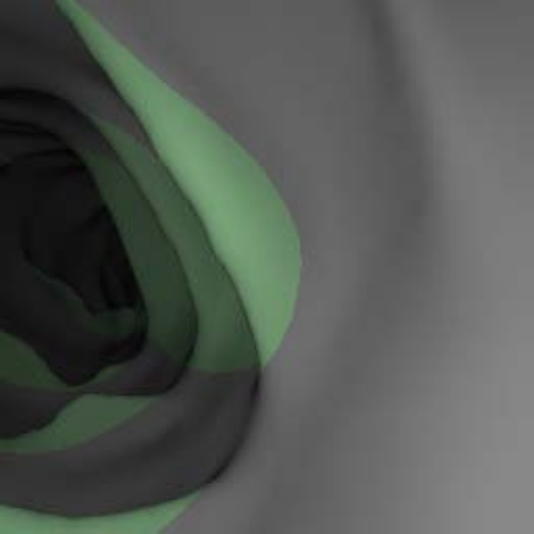}\\
(a) & (b)  & (c) & (d) \\ 
\end{tabular}
\ 
\begin{tabular}{ccc}
\small OC Input & \small XDCycleGAN & \small Ours \\
\includegraphics[width=0.11\textwidth]{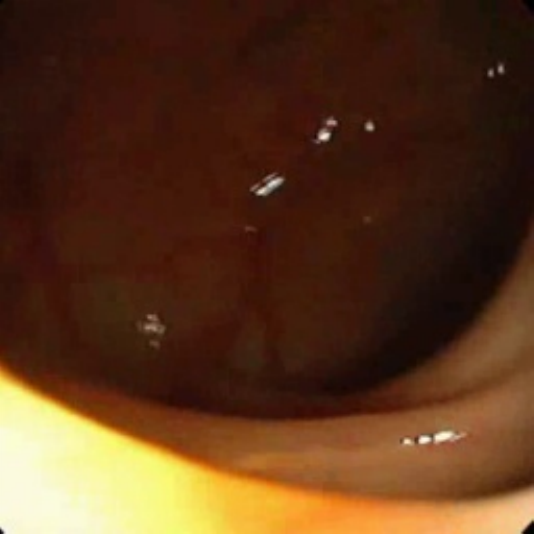}&
\includegraphics[width=0.11\textwidth]{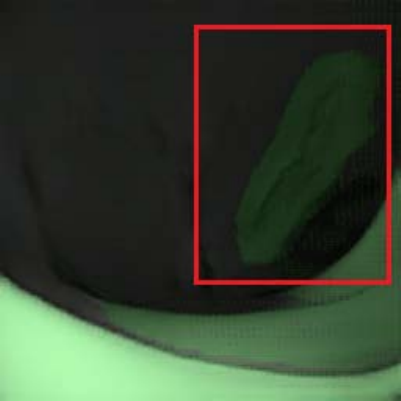}&
\includegraphics[width=0.11\textwidth]{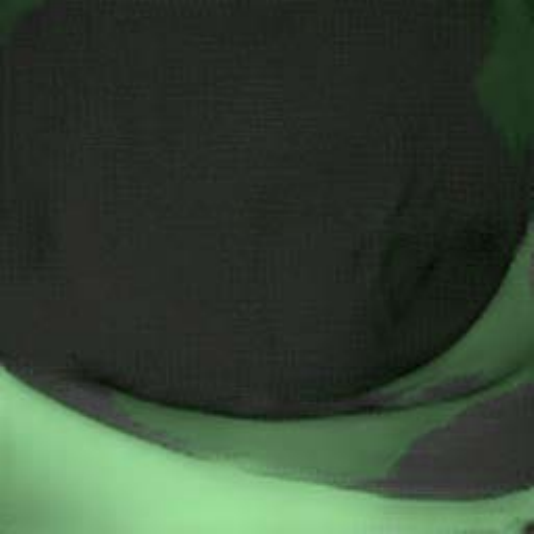}\\
\includegraphics[width=0.11\textwidth]{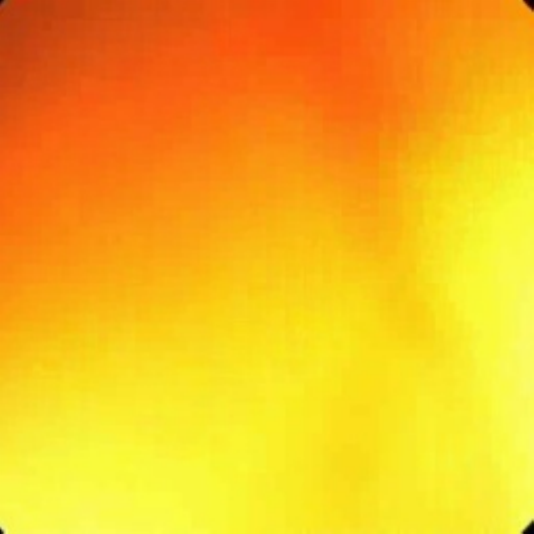}&
\includegraphics[width=0.11\textwidth]{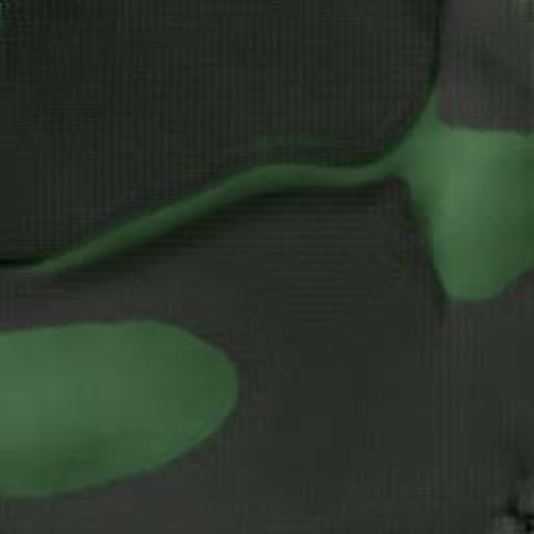}&
\includegraphics[width=0.11\textwidth]{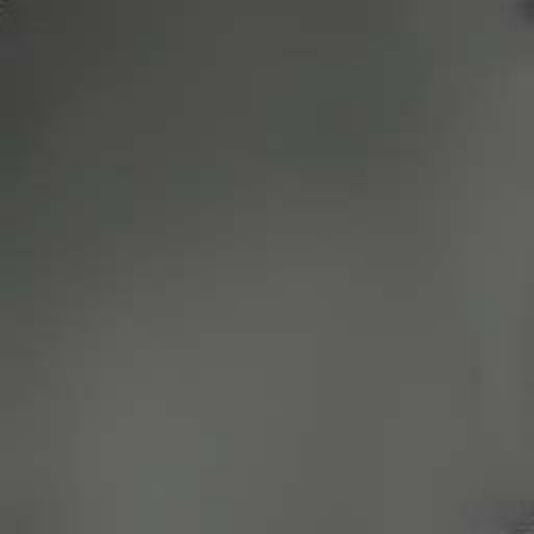}\\
\end{tabular}
\caption{(a) Pictorial representation of a colon illustrating missed colon surface (green outline) when the endoscope (black line) traverses from rectum (R) to cecum (C) and back. (b) A VC rendering for a mesh reconstructed from a CT scan. (c) The missing surfaces (green regions) can be rendered by casting rays from virtual camera positions and marking mesh faces not directly intersecting the rays. (d) Missing colon surface visualization with a virtual colonoscopy rendering is used for training the model. At the bottom are examples of XDCycleGAN \cite{mathew2020augmenting}, proposed for scale-consistent depth map inference across colonoscopy video frames, overfitting for missing surface inference task due to sparse training data, as shown with the red bounding box and the predicted missed structures (bottom) when the camera is occluded.}
\label{fig:data}
\end{center}
\end{figure}

More than 15 million colonoscopies are performed in the US every year \cite{joseph2016colorectal,seeff2004many}. During these procedures, 22-28\% of polyps and 20-24\% adenomas are missed \cite{leufkens2012factors}. There are no commercial or automated tools available to assist endoscopists in gauging the amount of colon surface missed during optical colonoscopy (OC) procedures. 
The main culprit in substandard coverage during colonoscopy are the sharp bends and haustral folds, as depicted in Figure \ref{fig:data}a. Even though the endoscope tip can be flexed to look behind folds and sharp bends, beginner or tired endoscopists might not use this option wisely and may have a high miss rate. This high miss rate can be reduced if endoscopists have a visualization tool to identify and investigate areas occluded by haustral folds. 

Recently, Ma et al. \cite{ma2019real} and Freedman et al. \cite{freedman2020detecting} have presented approaches to quantify colon surface coverage. Ma et al. \cite{ma2019real} reconstruct 3D mesh from contiguous chunk of colonoscopy video frames using training data generated from shape-from-motion. The missed surface is visualized as holes in the reconstructed mesh. The method, however, assumes cylindrical topology (endoluminal) views, smooth camera movements and masked-out specular reflection, making the method less practical in general colonoscopy scenarios. In contrast, Freedman et al. \cite{freedman2020detecting} have used a deep learning approach to estimate percentage coverage value directly for given colonoscopy video segments but do not provide any means for visualizing the missed colon surface.

In this paper, we present a deep learning model for realtime visualization of missed colon surfaces directly on the colonoscopy video frames without doing any prior offline 3D reconstruction using contiguous sets of frames. 
Specifically, we make use of prior 3D reconstructed virtual colonoscopy (VC) \cite{hong1997virtual,pickhardt2003computed} data, created from a computed tomography (CT) scan, to produce training data for missing surface visualization (Figure \ref{fig:data}b--d). This is used in conjunction with OC data for the same patient to drive an unpaired image-to-image translation with a modified lossy CycleGAN \cite{mathew2020augmenting} and a new enforced shared OC and VC latent space representation. The lossy CycleGAN \cite{mathew2020augmenting} by itself overfits due to the sparse training data for the missing surface task (most OC frames have no or few missing surface green pixels as opposed to the dense depth maps for which the lossy CycleGAN was originally proposed) and can easily hallucinate structures which do not exist, as shown in Figure \ref{fig:data}. Adding a shared latent space forces the network to preserve structures (and avoid hallucination) when translating between domains. With added Gaussian noise, we also show that the same framework with shared latent space representations can be used to generate realistic one-to-many mappings from VC to OC and OC to OC for augmenting OC datasets in computer-aided detection and classification pipelines.

In summary, the contributions of this paper are as follows:
\begin{itemize}
    \item We are the first to present a model to visualize missing surface regions for individual colonoscopy frames in realtime.
    \item We introduce shared OC and VC latent space representations to get more consistent geometry for missing surface inference task.
    \item Using additional Gaussian noise input, the model can also generate realistic OC images (one-to-many mapping) with different specular reflections, lighting and texture for a given OC or VC frame.
\end{itemize}

\section{Data}
The OC and VC datasets were acquired for 10 patients at Stony Brook University Hospital. 2000 images from 5 patients were used for training, while 800 images from 2 patients were used for validation and 1200 images from 3 patients were used for testing. Even though VC and OC are captured for the same patient, there is no ground truth since the shape of the colon differs considerably between the two procedures. The borders in the OC images were cropped to exclude the fisheye lens artifacts. 3D meshes were reconstructed from CT scans in VC, using a pipeline similar to Nadeem and Kaufman \cite{nadeem2016computer}.

In order to create training data for per-frame missing surface visualization, the opacity of the 3D colon mesh is lowered such that the more opaque regions indicate the missed surfaces, which are colored green in Figure \ref{fig:data}c. The per-frame missing surface data is generated through Blender and example videos are provided\footnote{Supplementary Video: \url{https://youtu.be/x1-wwCiYeC0}}. Figure \ref{fig:data} shows a typical colon anatomy along with the haustral folds and the pictorial representation of a missed surface for a certain endoscope camera position. To aid the model with the image-to-image domain translation task, we added the missing surface information in green channel on top of the VC rendering of the colon (Figure \ref{fig:data}d).

\section{Method}

\begin{figure}[t!]
\begin{center}
\includegraphics[width=0.48\textwidth]{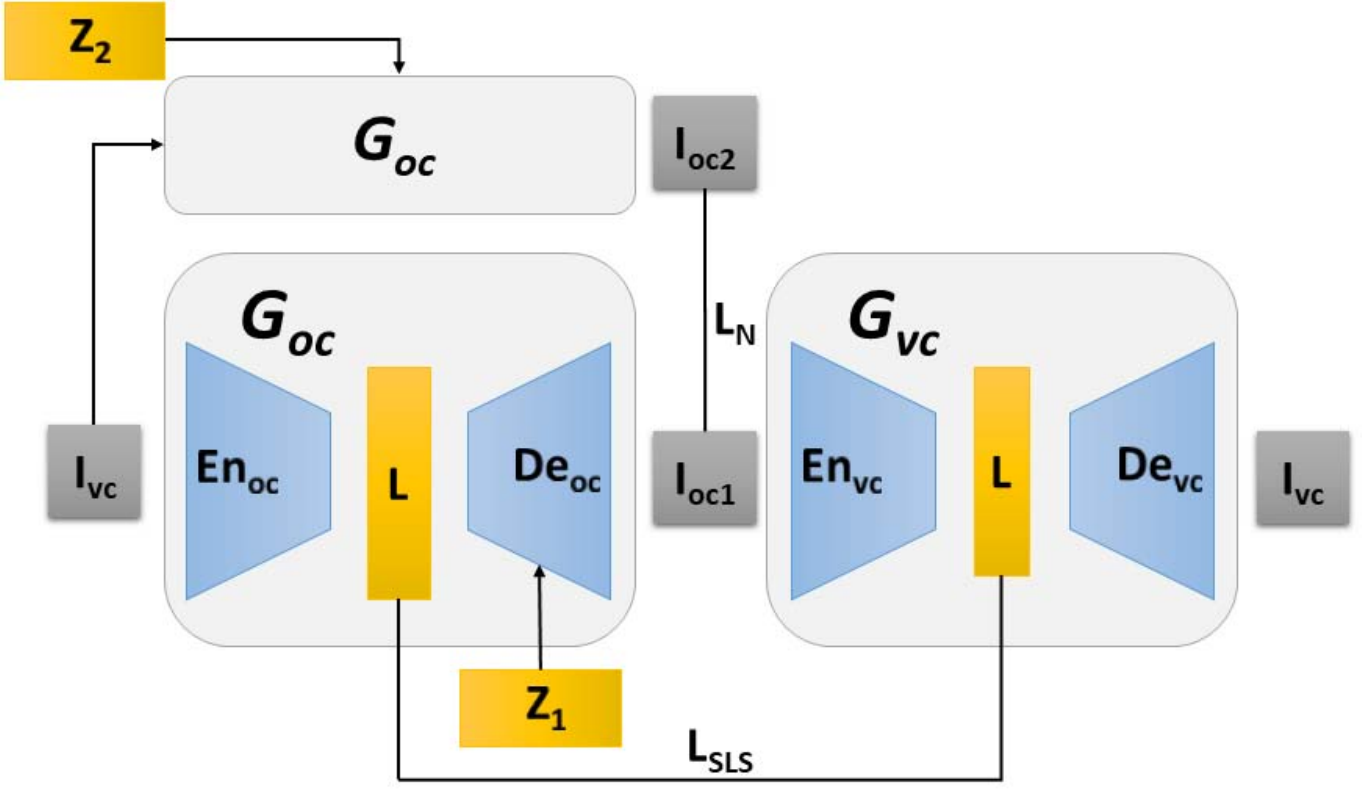}\\
\caption{A VC image, $I_{vc}$, is brought into latent space with $En_{oc}$, and brought to the OC domain through $De_{oc}$. The OC image, $I_{oc}$ returns to latent space through $En_{vc}$. The latent vector produced by $En_{oc}$ and $En_{vc}$ should be the same, since it stores the same geometric information. This is enforced by $\mathcal{L}_{SLS}$. To generate different OC images, $De_{oc}$ takes noise input, $Z$, to create the additional OC information. In order to ensure the network uses the noise, we apply a loss, $\mathcal{L}_{N}$, between OC images with different noise vectors.}
\label{fig:network}
\end{center}
\end{figure}

The presented approach focuses on training two generator networks, $G_{oc}$ and $G_{vc}$ and two discriminator networks. To train these networks, an objective function $\mathcal{L}$ composed of three parts is used. The first part, $\mathcal{L}_{trans}$ focuses on learning a proper image-to-image translation. $\mathcal{L}_{adv}$ produces realistic images, and $\mathcal{L}_{noise}$ helps the network utilize noise input for OC image generation:
\begin{equation}
\mathcal{L} = \mathcal{L}_{trans} +\mathcal{L}_{adv} + \mathcal{L}_{noise}
\end{equation}

In order to learn the image-to-image translation, a cycle consistency loss, $\mathcal{L}_{cyc}$, and an extended cycle consistency loss, $\mathcal{L}_{excyc}$ \cite{mathew2020augmenting} are used. $\mathcal{L}_{excyc}$ allows for a one-to-many translation by making comparisons in a common domain. The common domain between OC and VC is VC, since OC has additional textures, lighting and specular reflections. The cycle consistency and extended cycle consistency losses are as follows:

\begin{equation}
    \mathcal{L}_{cyc}(G_a,G_b,A) = \mathds{E}_{y \backsim p(A)} \|y - G_a(G_b(y))\|_1
\end{equation}

\begin{equation}
    \mathcal{L}_{ excyc}(G_a,G_b,A) = \mathds{E}_{y \backsim p(A)} \|G_b(y) - G_b(G_a(G_b(y)))\|_1
\end{equation}
where $y \backsim p(A)$ is the data distribution of domain A and $\| \cdot \|_1$ represents the L1 norm. 

To make the translation more robust, we add a shared latent space loss. Each generator, $G_A$, is composed of an encoder, $En_A$, and decoder, $De_A$. 

The encoder brings the input image into latent space, while the decoder takes the latent space into the image domain. We propose that OC and VC share the same latent space, as the latent space stores geometric information which should be consistent between the two domains. This is depicted in Figure \ref{fig:network} and the following equation: 
\begin{equation}
\begin{split}
    \mathcal{L}_{SLS}(En_B,G_B,En_A,A) = \ &  \mathds{E}_{y \backsim p(A)}  \\& \| En_B(y) - En_A(G_B({y})))\|_1,
\end{split}
\end{equation}
Lastly, we add an identity loss, as described by Zhu et al. \cite{zhu2017unpaired}, in order to ensure consistent shading, when translating between the two domains. This is not done in the OC domain as it restricts the network from using the noise input properly.

\begin{equation}
\begin{split}
    \mathcal{L}_{trans} = \ & \lambda_c \big[ \mathcal{L}_{excyc}(G_{oc},G_{vc},I_{oc}) +  \mathcal{L}_{cyc}(G_{vc},G_{oc},I_{vc})\big]  \\ & +  \lambda_{SLS} \big[ \mathcal{L}_{SLS} (En_{oc},G_{oc},En_{vc},I_{vc}) \\ & +  \mathcal{L}_{SLS}(En_{vc},G_{vc},En_{oc},I_{oc}) \big]  \\ & +   \lambda_{iden}\mathcal{L}_{iden}(I_{vc}),
\end{split}
\end{equation}
In our experiments, we set $\lambda_c$ as 10, $\lambda_{SLS}$ as 1, and $\lambda_{iden}$ as 1.

Adversarial losses have shown success in producing realistic images. Specifically, we add a directional discriminator $D_{dir}$, to differentiate between the different directions of translation and an OC discriminator to restrain $G_{oc}(G_{vc}(I_{oc}))$, as described by \cite{mathew2020augmenting}. 
\begin{equation}
\begin{split}
    \mathcal{L}_{dir}(G_a,G_b,D,A,B) = \ & \mathds{E}_{y \backsim p(A)} \big[ \textrm{log} (D(y,G_b(y))\big] + \\ &  \mathds{E}_{x \backsim p(B)} \big[ \textrm{log} (1-D(G_a(x),x)\big]
\end{split}
\end{equation}

\begin{equation}
\begin{split}
    \mathcal{L}_{GAN}(G,D,A,B) = \ & \mathds{E}_{y \backsim p(A)} \big[ \textrm{log} (D(y)\big] + \\ & \mathds{E}_{x \backsim p(B)} \big[ \textrm{log} (1-D(G(x))\big]
\end{split}
\end{equation}

\noindent
These two losses compose the adversarial losses:
\begin{equation}
\begin{split}
    \mathcal{L}_{adv} = \ & \mathcal{L}_{dir}(G_{oc},G_{vc},D_{dir},I_{oc},I_{vc})  
    + \\ & \mathcal{L}_{GAN}(G_{oc},D_{oc},I_{oc},G_{oc}(G_{vc}(I_{oc})))
\end{split}
\end{equation}

We note that OC and VC share the same underlying geometric information, while OC has additional color, texture, and specular reflections. In order to reflect this additional information, we add a noise input to $De_{oc}$ to drive the one-to-many mapping between VC and OC. $\mathcal{L}_{noise}$ is used to ensure a minimum distance between images with different noises, otherwise the noise vector is ignored:
\begin{equation}
\begin{split}
    \mathcal{L}_{noise}(De,N,L) = \ &  \mathds{E}_{z_1,z_2 \backsim p(N), l \backsim p(L)} \\ &
    max(0, \|De(l,z_1)- De(l,z_2)\| - \alpha),
\end{split}
\end{equation}
$L$ is a latent space domain, $N$ is a noise domain, and $\alpha$ is a variable to determine how much the images should differ. We set $\alpha$ to 0.1 in our experiments and draw our noise input from a normal distribution. 

With $L_{noise}$, we can take a latent space variable along with various samples from the noise domain to produce OC images with different specular reflections, lighting, and texture. The latent space can be produced from both OC and VC images. With the addition of the noise input, we create a one-to-many mapping between VC-to-OC and OC-to-OC (Figure \ref{fig:noise}).

We follow the same generator architecture described in CycleGAN \cite{zhu2017unpaired} but instead of using 9 ResNet blocks, we use 10, 5 dedicated to encoder and the remaining 5 for decoder.

\begin{figure}[t!]
\begin{center}
\setlength{\tabcolsep}{0.5pt}
\begin{tabular}{cccccccc}
\rotatebox{90}{~\rlap{\footnotesize OC Input}}&
\includegraphics[width=0.07\textwidth]{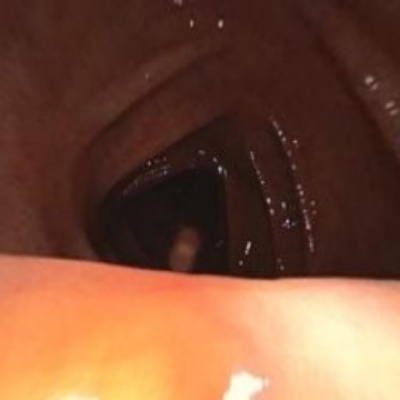}&
\includegraphics[width=0.07\textwidth]{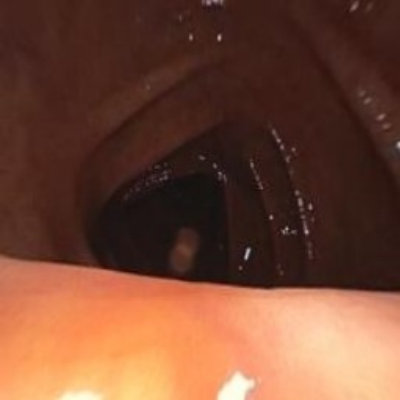}&
\includegraphics[width=0.07\textwidth]{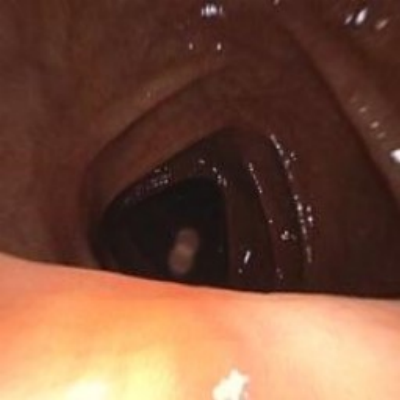}&
\ldots&
\includegraphics[width=0.07\textwidth]{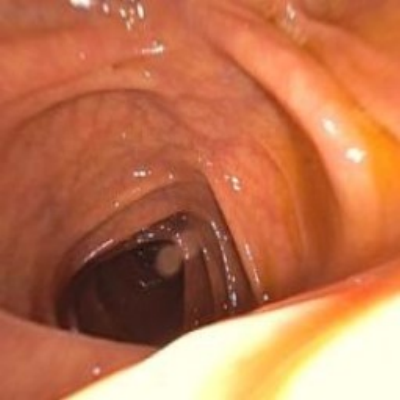}&
\includegraphics[width=0.07\textwidth]{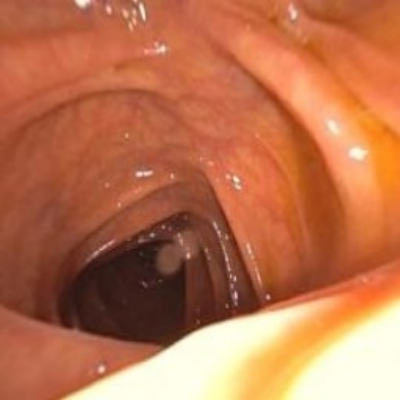}&
\includegraphics[width=0.07\textwidth]{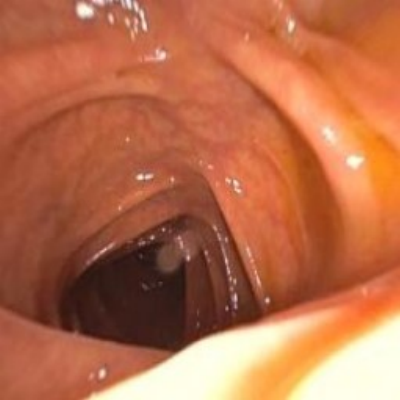}\\

\rotatebox{90}{~~~~\rlap{\footnotesize Ours}}&
\includegraphics[width=0.07\textwidth]{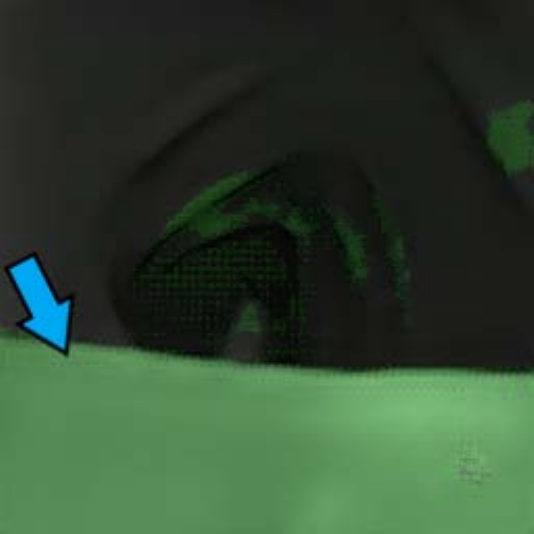}&
\includegraphics[width=0.07\textwidth]{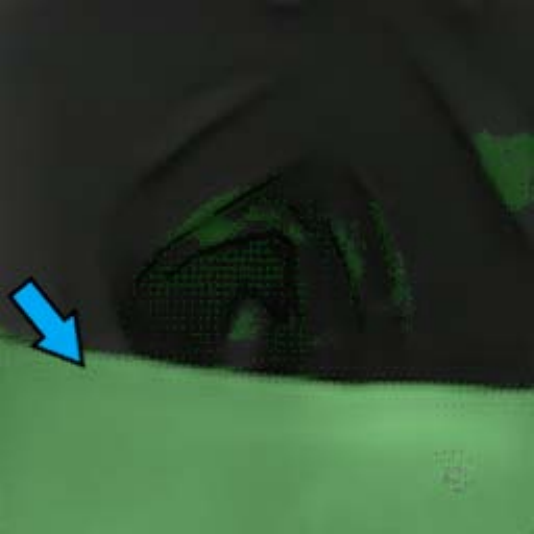}&
\includegraphics[width=0.07\textwidth]{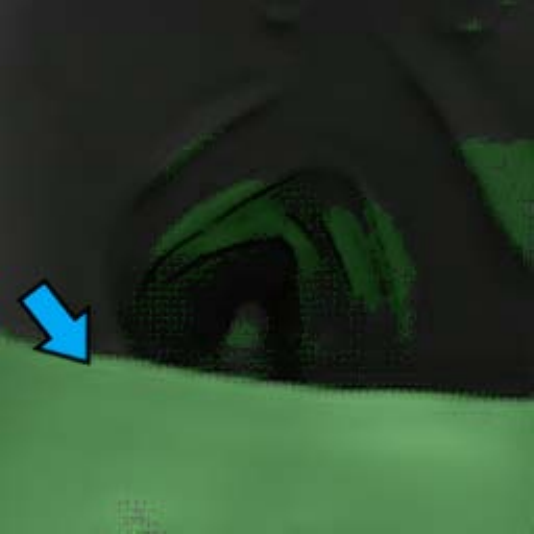}&
\ldots&
\includegraphics[width=0.07\textwidth]{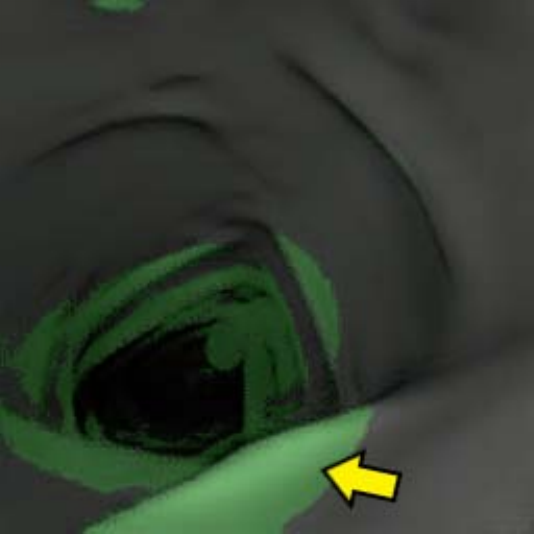}&
\includegraphics[width=0.07\textwidth]{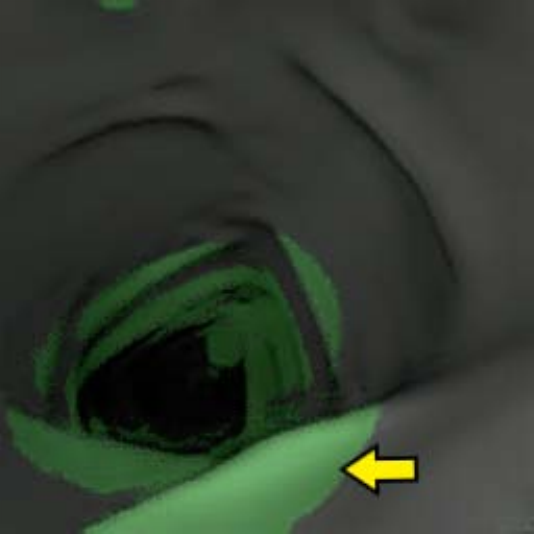}&
\includegraphics[width=0.07\textwidth]{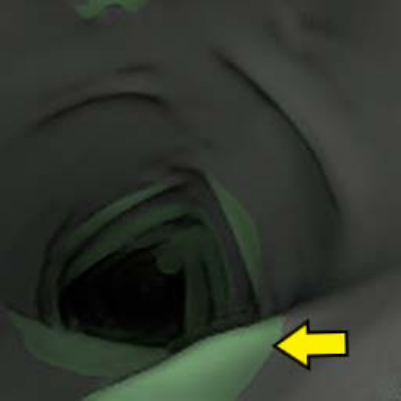}\\

\rotatebox{90}{~\rlap{\footnotesize Ma et al.}}&
\includegraphics[width=0.07\textwidth]{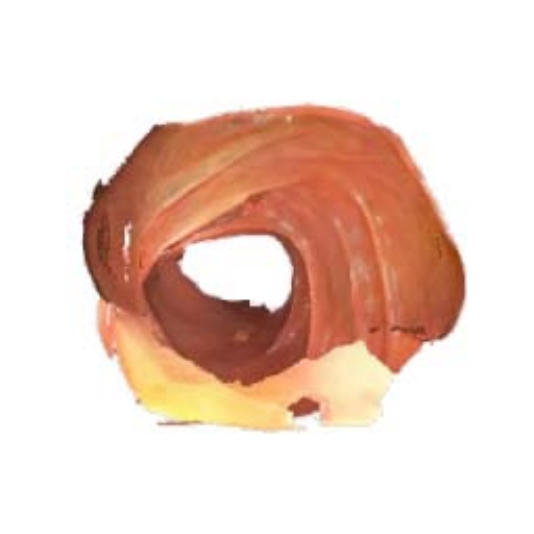}&
\includegraphics[width=0.07\textwidth]{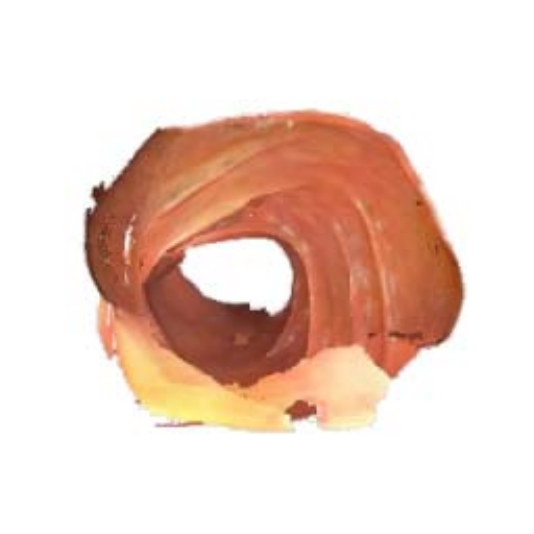}&
\includegraphics[width=0.07\textwidth]{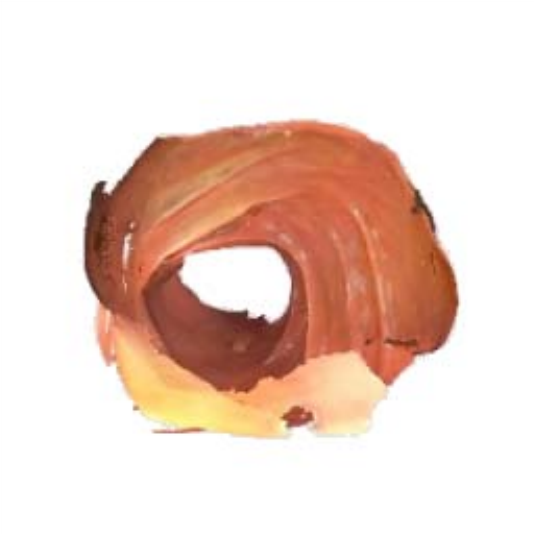}&
\ldots&
\includegraphics[width=0.07\textwidth]{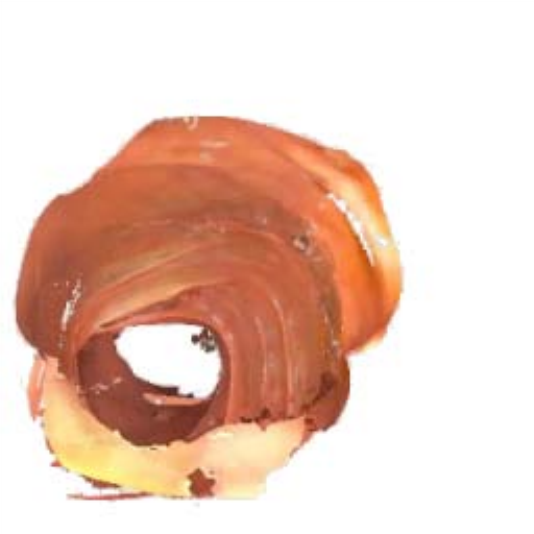}&
\includegraphics[width=0.07\textwidth]{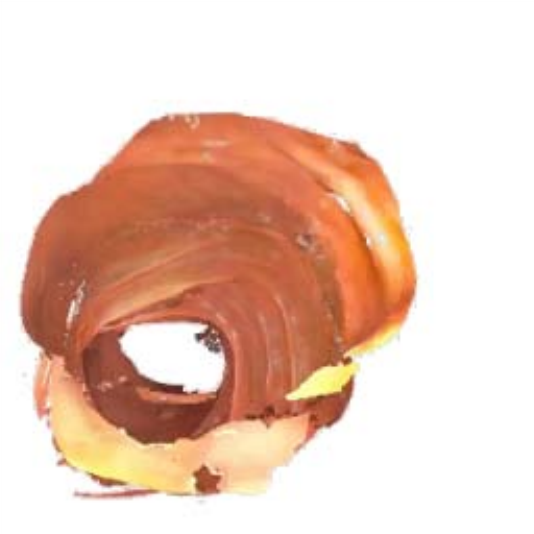}&
\includegraphics[width=0.07\textwidth]{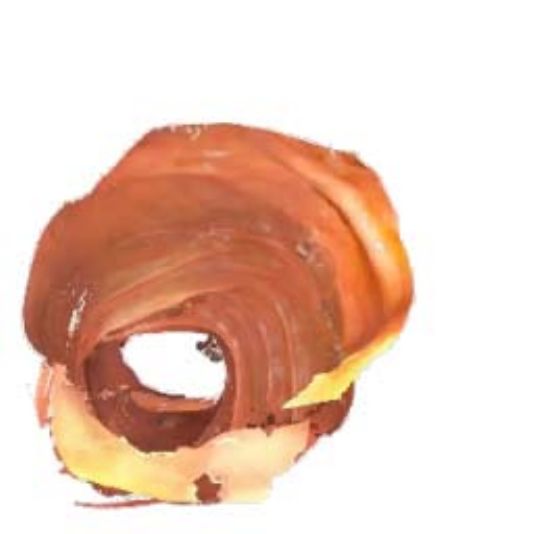}\\

\rotatebox{90}{~~~~~~~\rlap{\footnotesize Ma et al.}}&
\multicolumn{7}{c}{\includegraphics[width=0.35\textwidth]{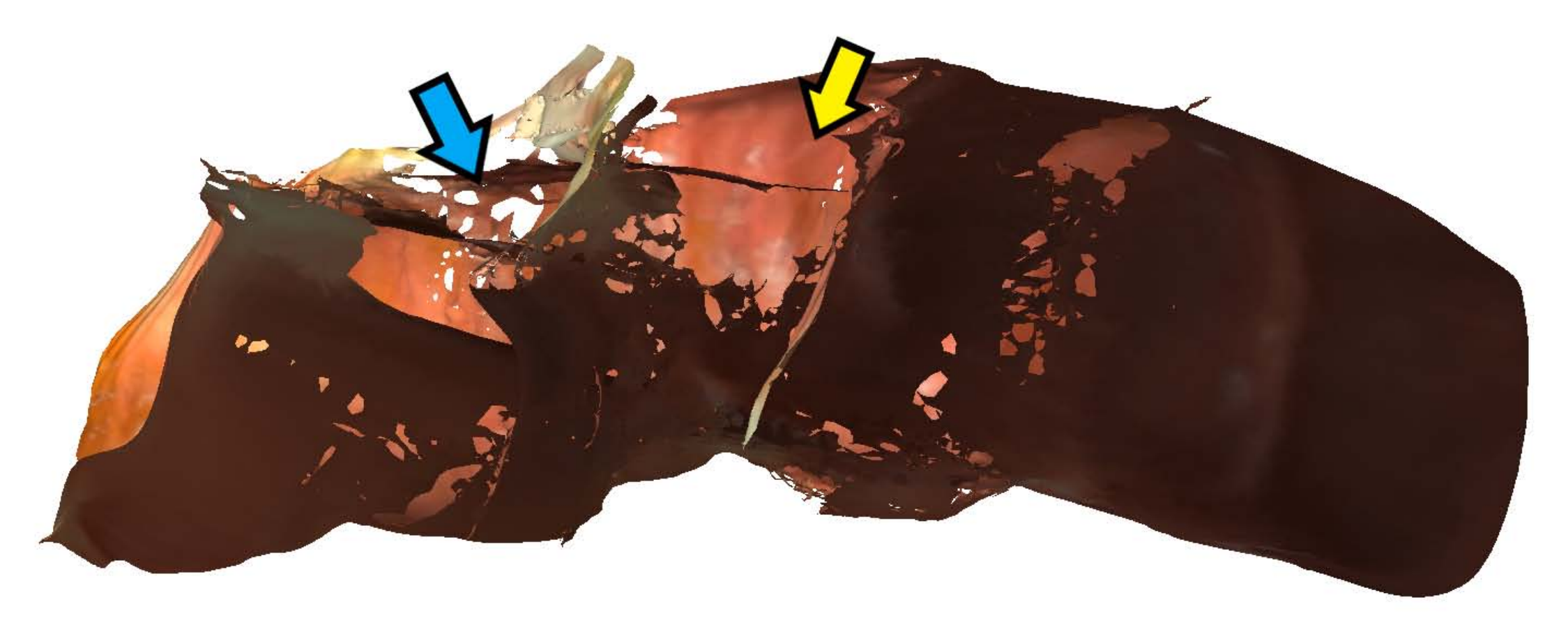}}\\

\rotatebox{90}{\rlap{\footnotesize OC Input}}&
\includegraphics[width=0.07\textwidth]{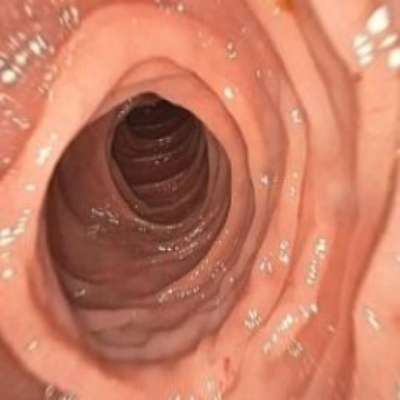}&
\includegraphics[width=0.07\textwidth]{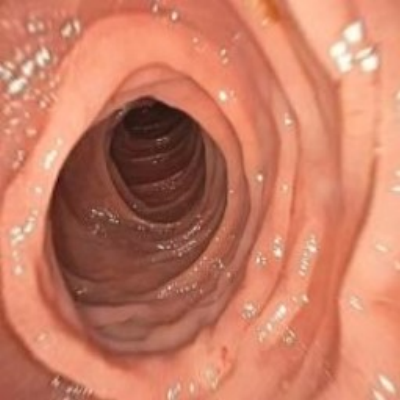}&
\includegraphics[width=0.07\textwidth]{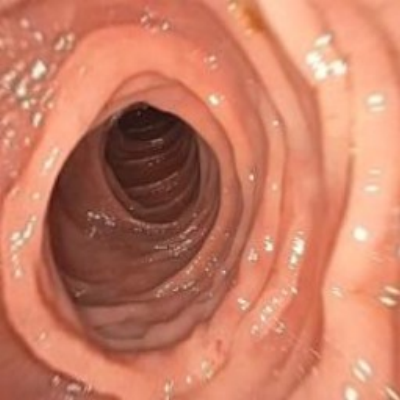}&
\ldots&
\includegraphics[width=0.07\textwidth]{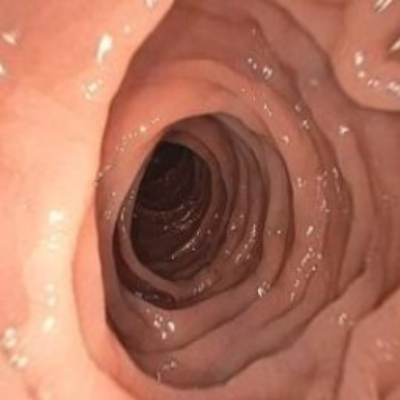}&
\includegraphics[width=0.07\textwidth]{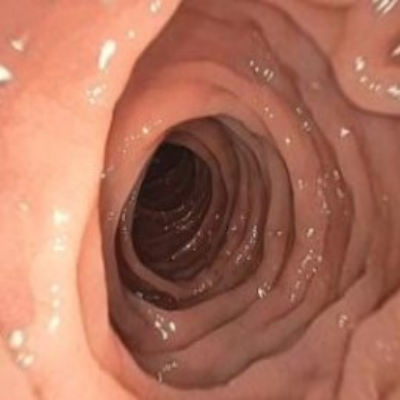}&
\includegraphics[width=0.07\textwidth]{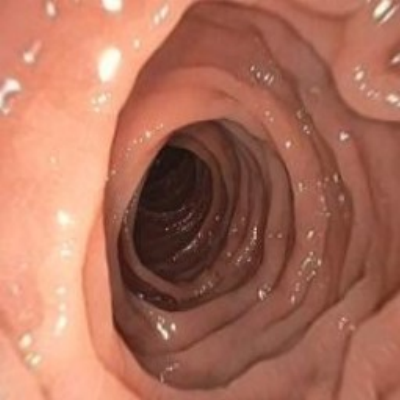}\\

\rotatebox{90}{~~~~\rlap{\footnotesize Ours}}&
\includegraphics[width=0.07\textwidth]{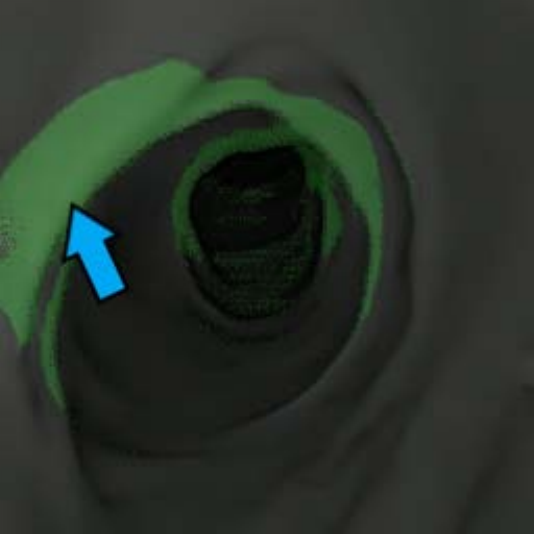}&
\includegraphics[width=0.07\textwidth]{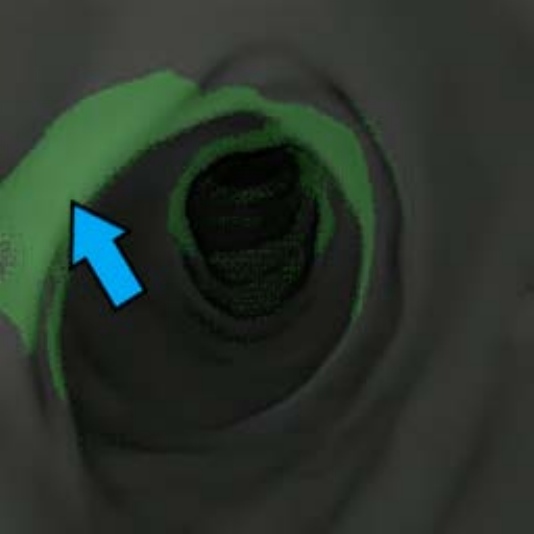}&
\includegraphics[width=0.07\textwidth]{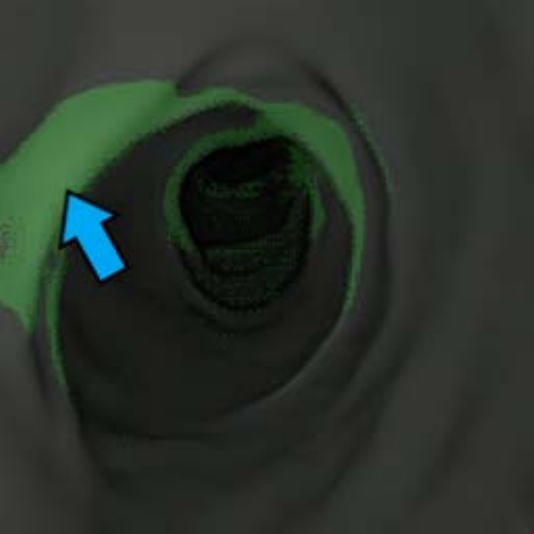}&
\ldots&
\includegraphics[width=0.07\textwidth]{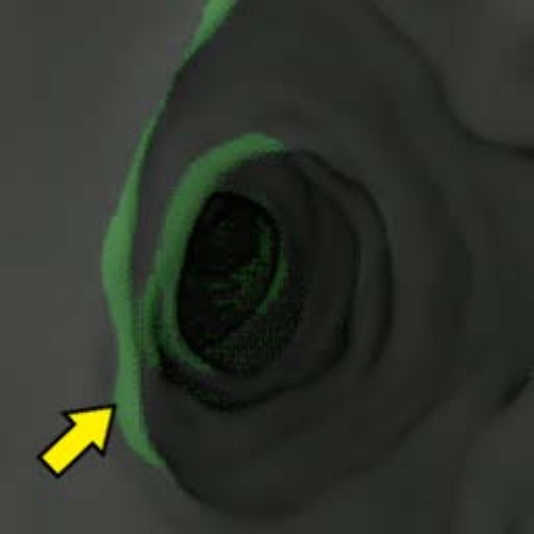}&
\includegraphics[width=0.07\textwidth]{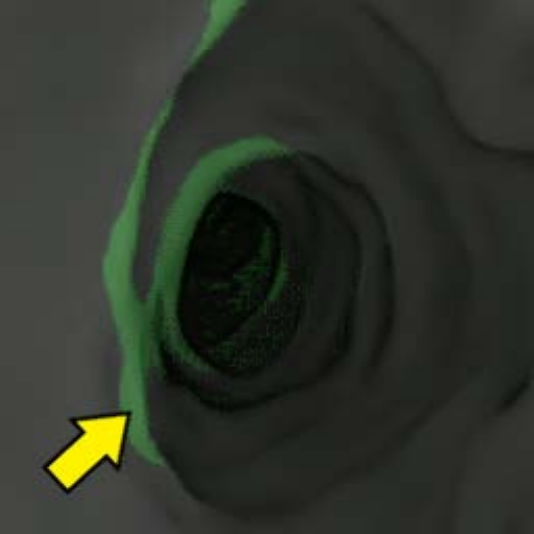}&
\includegraphics[width=0.07\textwidth]{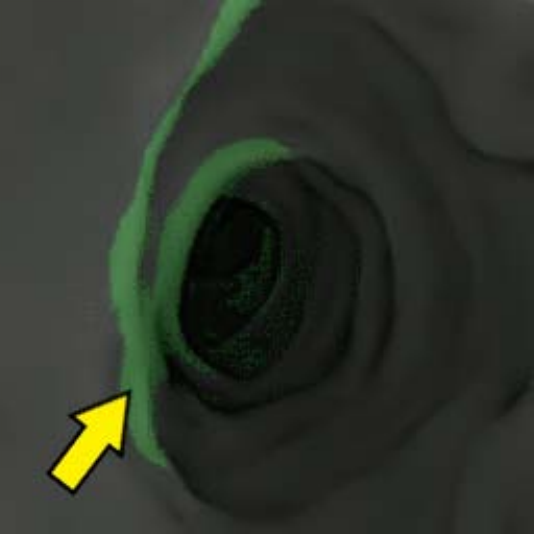}\\

\rotatebox{90}{\rlap{\footnotesize Ma et al.}}&
\includegraphics[width=0.07\textwidth]{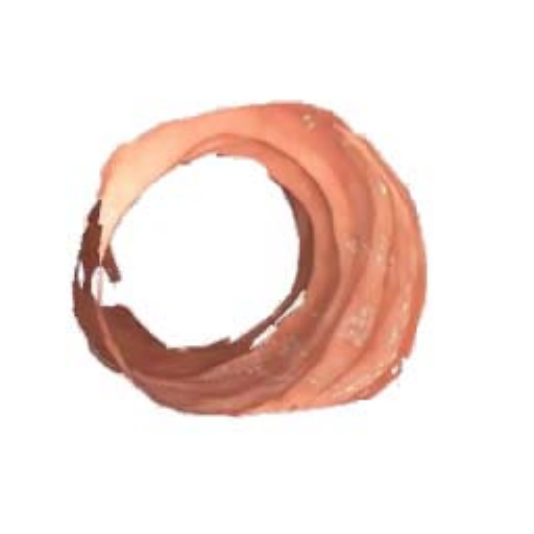}&
\includegraphics[width=0.07\textwidth]{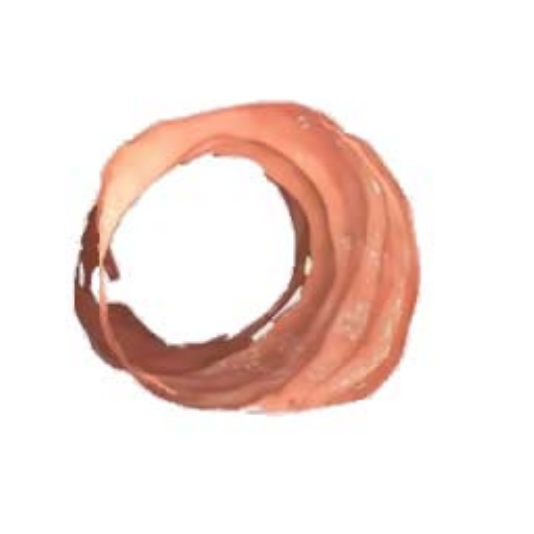}&
\includegraphics[width=0.07\textwidth]{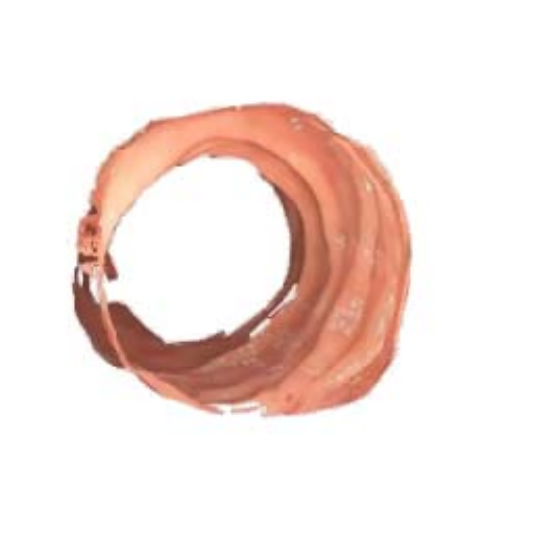}&
\ldots&
\includegraphics[width=0.07\textwidth]{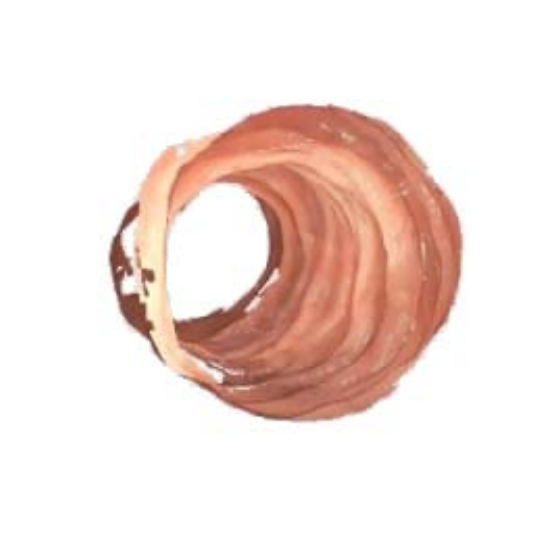}&
\includegraphics[width=0.07\textwidth]{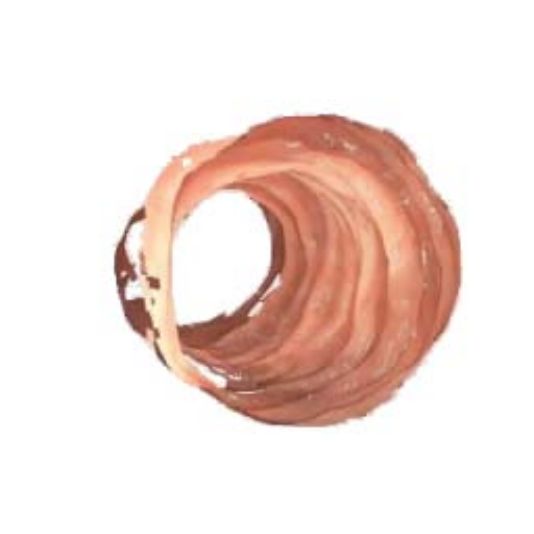}&
\includegraphics[width=0.07\textwidth]{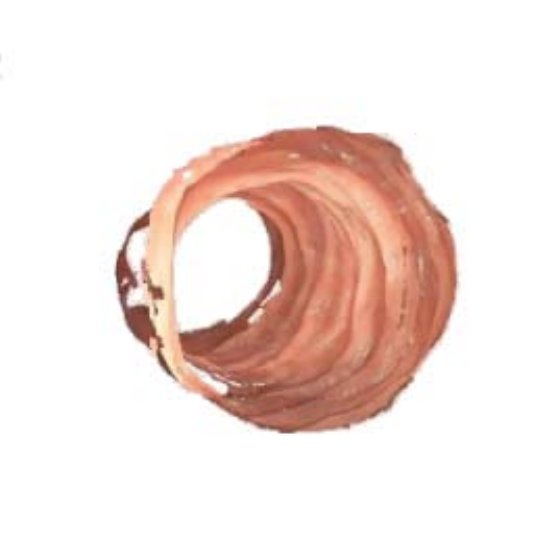}\\

\rotatebox{90}{\rlap{~~~~~~~\footnotesize Ma et al.}}&
\multicolumn{7}{c}{\includegraphics[width=0.35\textwidth]{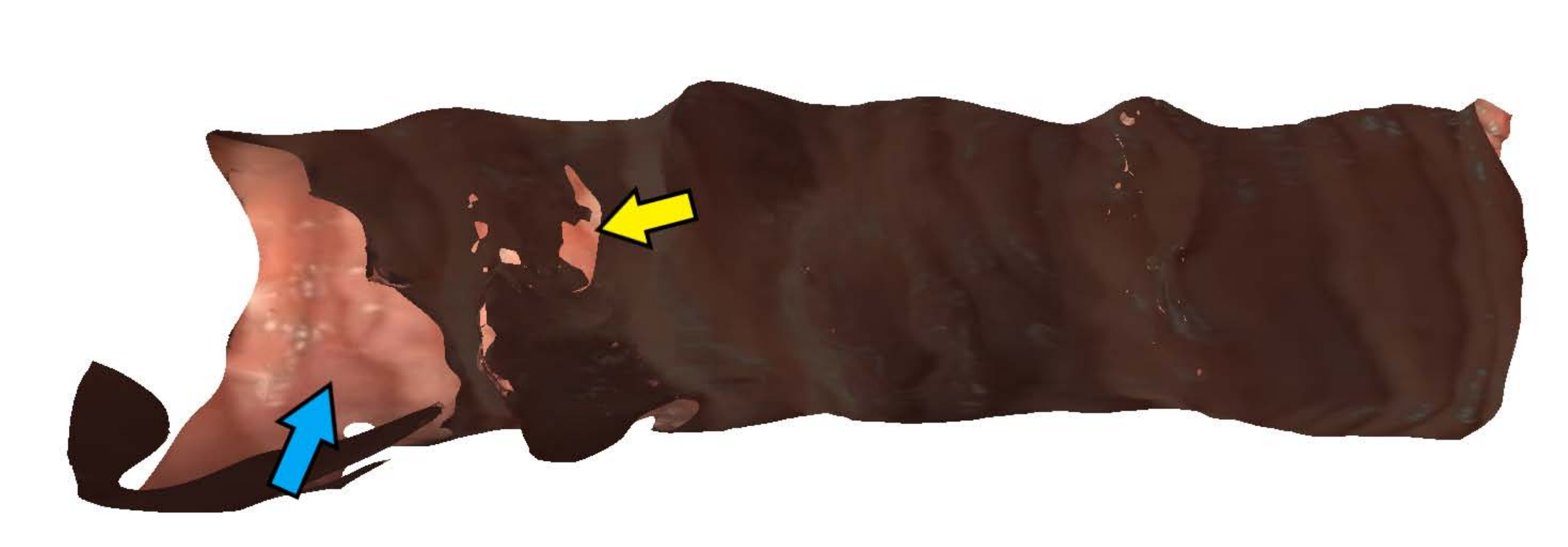}}\\

\hline\\

\rotatebox{90}{\rlap{\scriptsize Textured VC}}&
\includegraphics[width=0.07\textwidth]{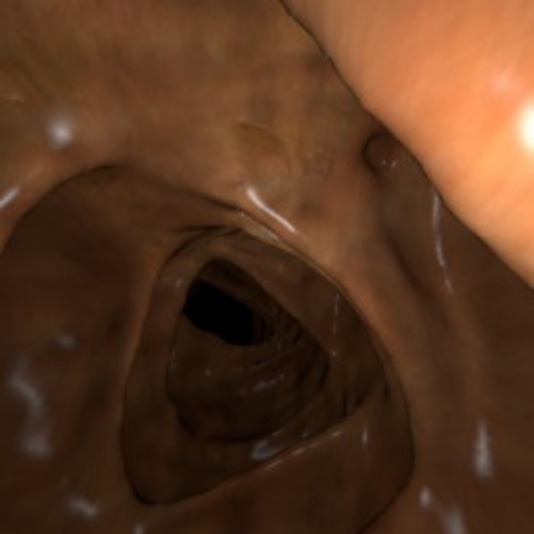}&
\includegraphics[width=0.07\textwidth]{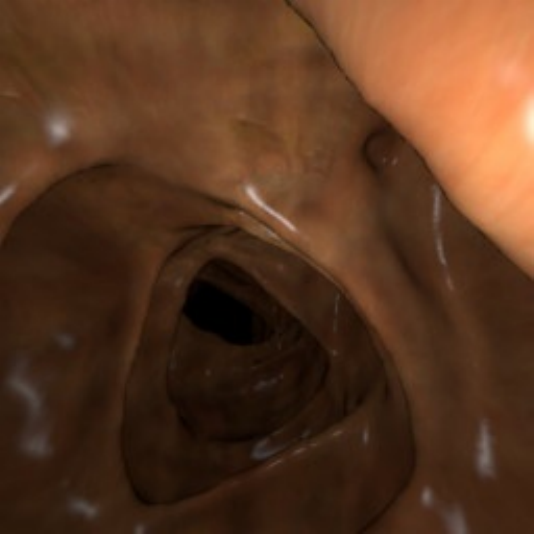}&
\includegraphics[width=0.07\textwidth]{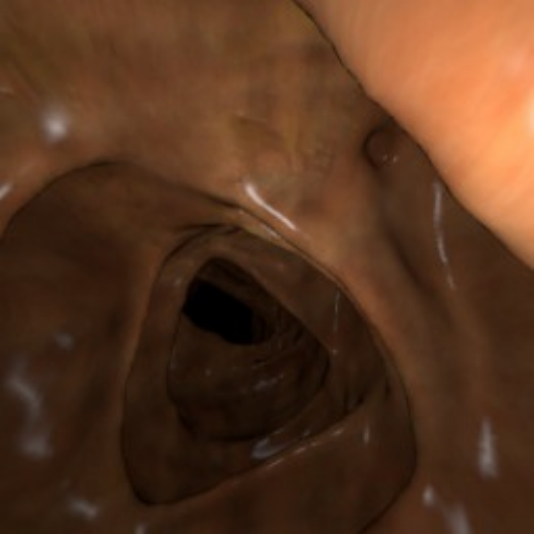}&
\ldots&
\includegraphics[width=0.07\textwidth]{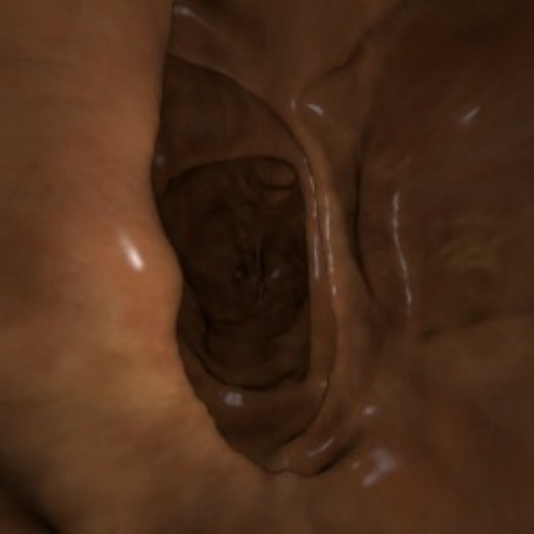}&
\includegraphics[width=0.07\textwidth]{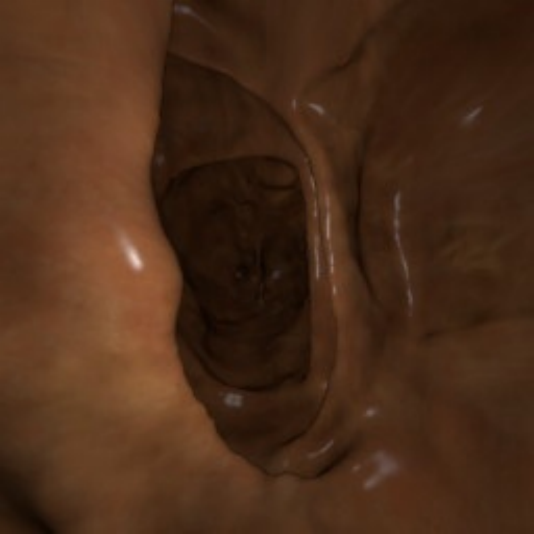}&
\includegraphics[width=0.07\textwidth]{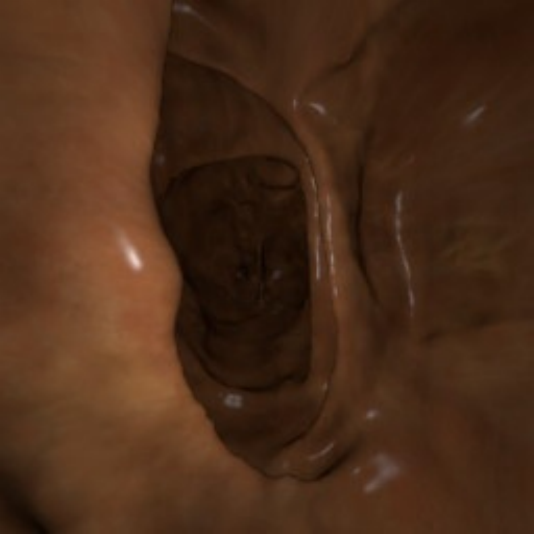}\\

\rotatebox{90}{~~~~\rlap{\footnotesize Ours}}&
\includegraphics[width=0.07\textwidth]{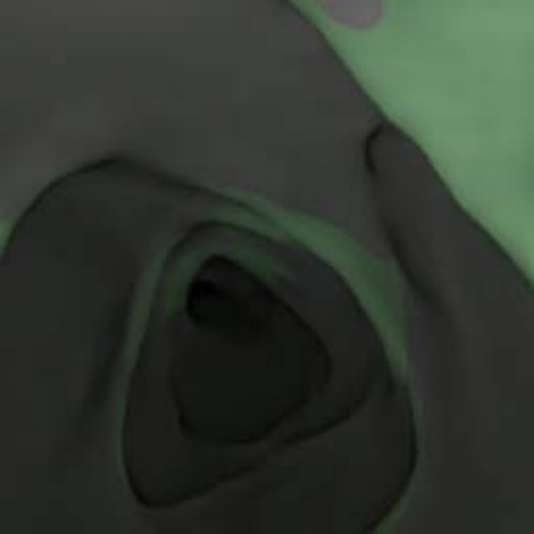}& 
\includegraphics[width=0.07\textwidth]{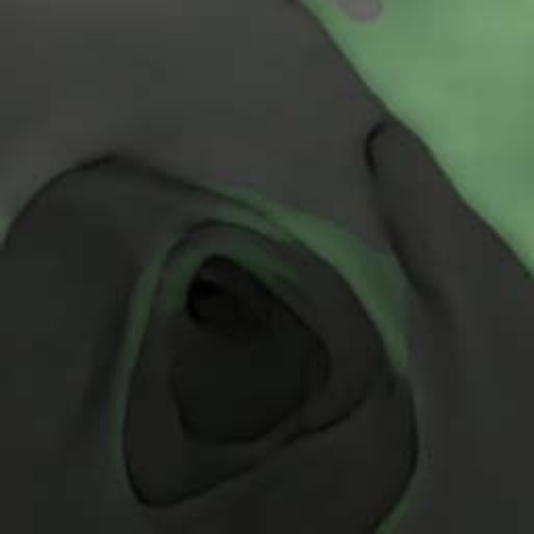}& 
\includegraphics[width=0.07\textwidth]{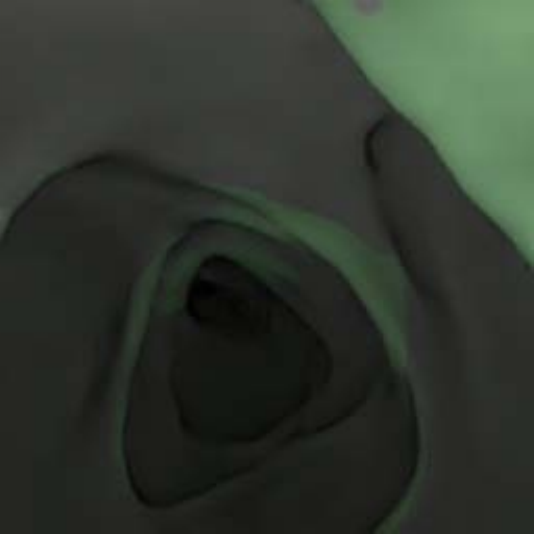}& 
\ldots&
\includegraphics[width=0.07\textwidth]{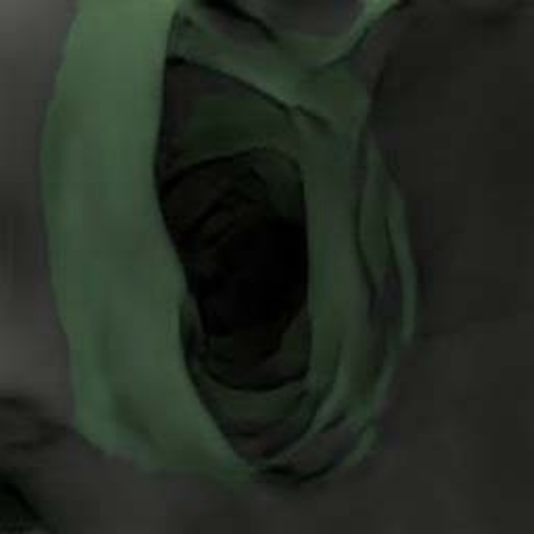}&
\includegraphics[width=0.07\textwidth]{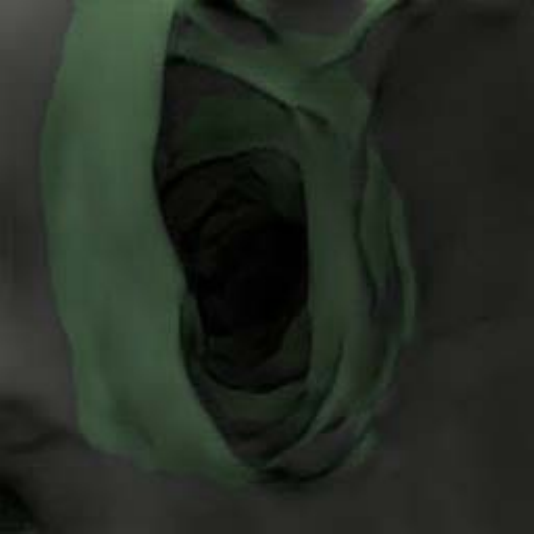}&
\includegraphics[width=0.07\textwidth]{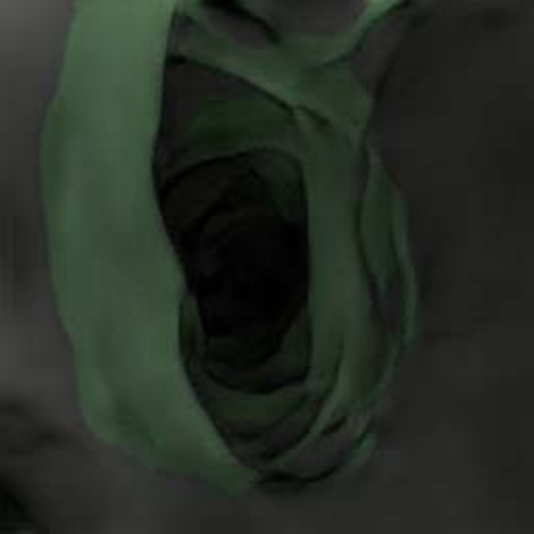}\\

\rotatebox{90}{\rlap{\scriptsize Ground Truth}}&
\includegraphics[width=0.07\textwidth]{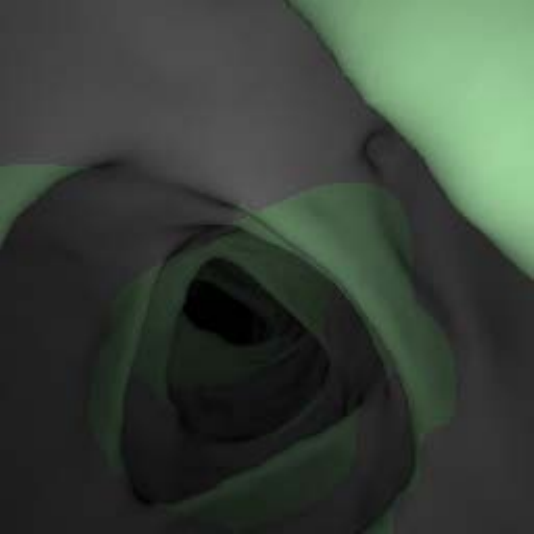}&
\includegraphics[width=0.07\textwidth]{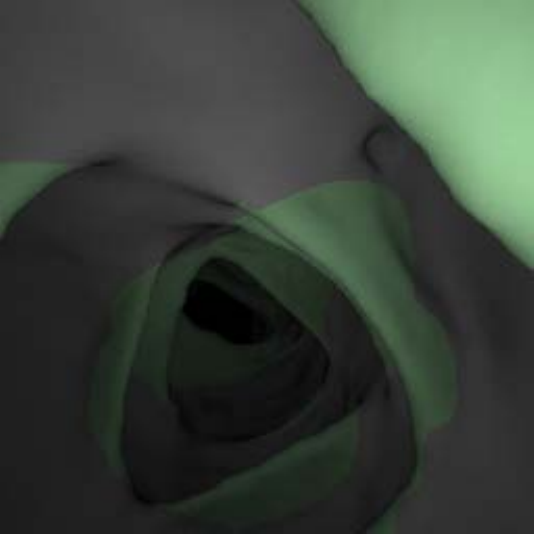}&
\includegraphics[width=0.07\textwidth]{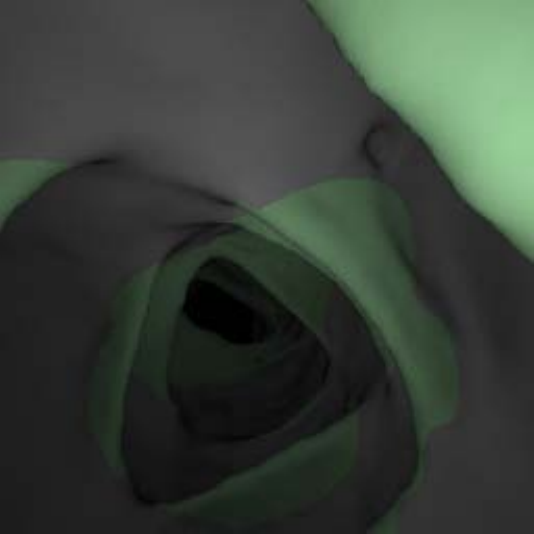}&
\ldots&
\includegraphics[width=0.07\textwidth]{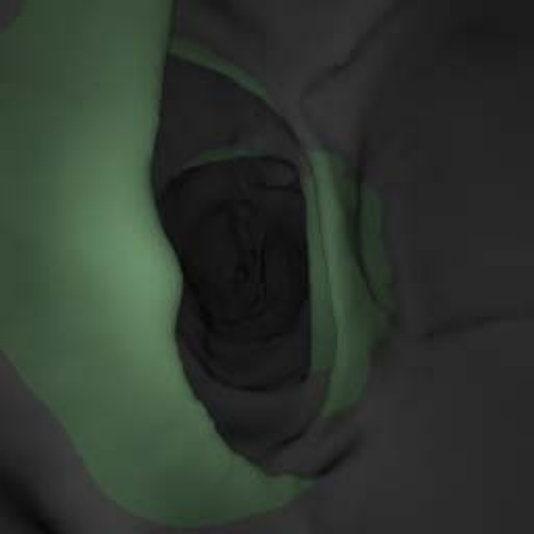}&
\includegraphics[width=0.07\textwidth]{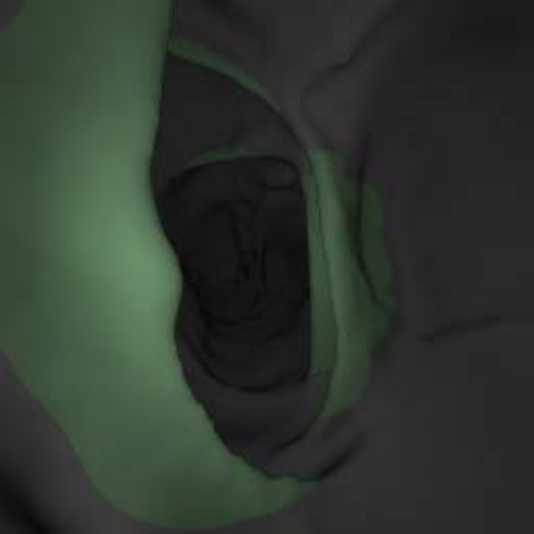}&
\includegraphics[width=0.07\textwidth]{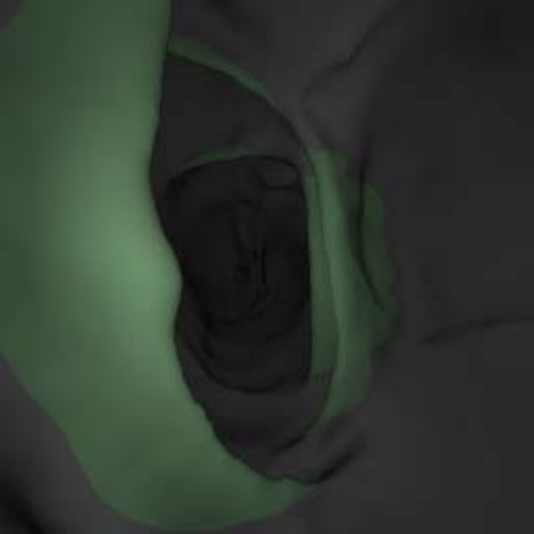}\\
\end{tabular}
\caption{The top portion shows results on video sequences from Ma et al. \cite{ma2019real}. We indicate the missing regions predicted by our model on the meshes reconstructed by their pipeline using blue and yellow arrows. The missing regions are visualized as holes on the reconstructed meshes. The bottom portion shows our results on textured VC input along with the ground truth.}
\vspace{-1em}
\label{fig:unc}
\end{center}
\end{figure}

\section{Results}

In Figure \ref{fig:unc}, we show our results on OC frames from Ma et al. \cite{ma2019real}. Ma et al. can produce 3D meshes from given OC video segments and can visualize the missing surfaces as holes in the reconstructed mesh. We highlight some holes in their reconstructed mesh that are detected by our model. Our results are similar for neighboring frames, without the use of any temporal connections for smoothing. 

There is no ground truth for missed surfaces in OC data. To make a quantitative analysis, we texture a VC colon to create ground truth missed surface data. This is done by taking texture from OC frames, and mapping them on the VC colon mesh. Our method achieved an average per-pixel accuracy score of 81\% and a Dice coefficient of .667 for the textured VC frames, despite the fact that surface area occluded by deep folds is difficult to predict in a single frame without additional information such as colon topology. Per-pixel accuracy is computed after converting the images into  binary images based areas classified as missed:
\begin{equation}
    Acc = \frac{TP+TN}{d},
\end{equation}
where $TP$ and $TN$ are the number of true positive and true negative pixels and $d$ is the number of pixels in the image. The textured VC results are shown in the last three rows of Figure \ref{fig:unc}. Complete videos are included in the supplement\footnote{Supplementary Video: \url{https://youtu.be/x1-wwCiYeC0}}.

\begin{figure}[t!]
\begin{center}
\setlength{\tabcolsep}{1.5pt}
\begin{tabular}{c|ccccc}

\includegraphics[width=0.09\textwidth]{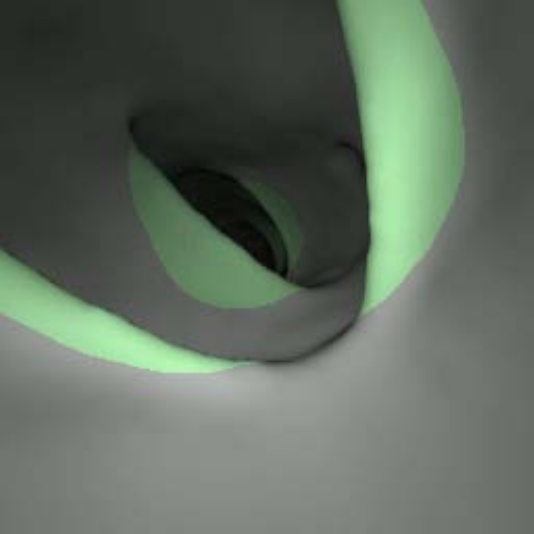}&
\includegraphics[width=0.09\textwidth]{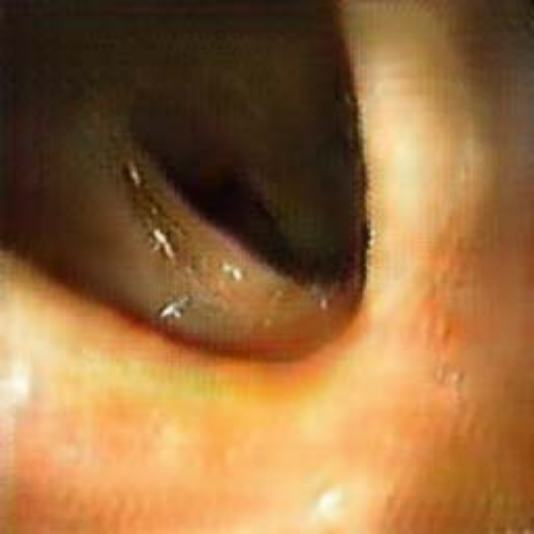}&
\includegraphics[width=0.09\textwidth]{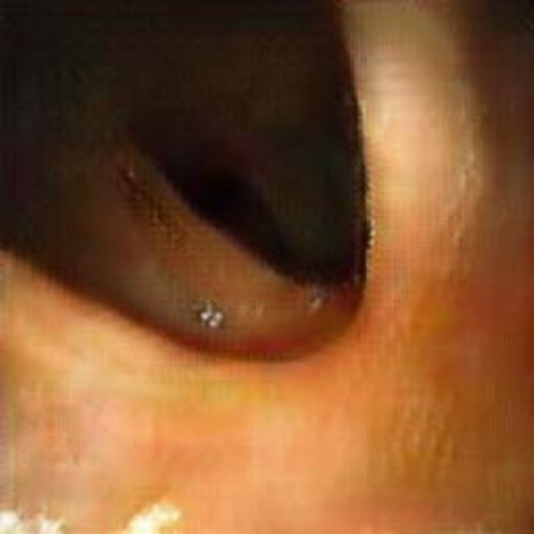}&
\includegraphics[width=0.09\textwidth]{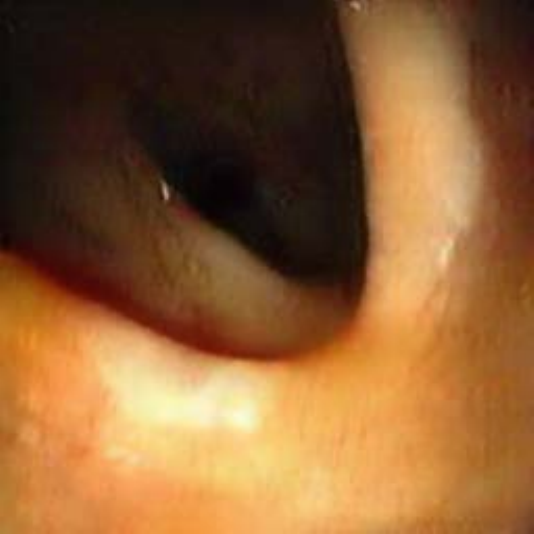}&
\includegraphics[width=0.09\textwidth]{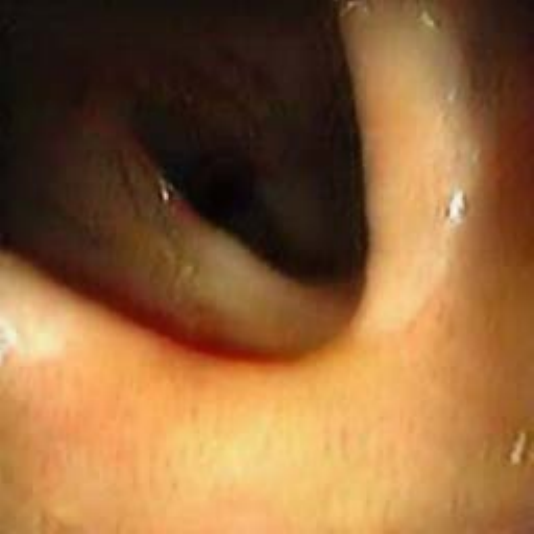}\\
\includegraphics[width=0.09\textwidth]{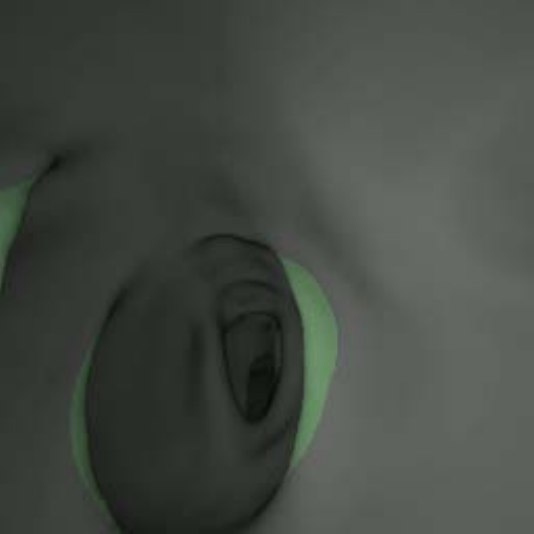}&
\includegraphics[width=0.09\textwidth]{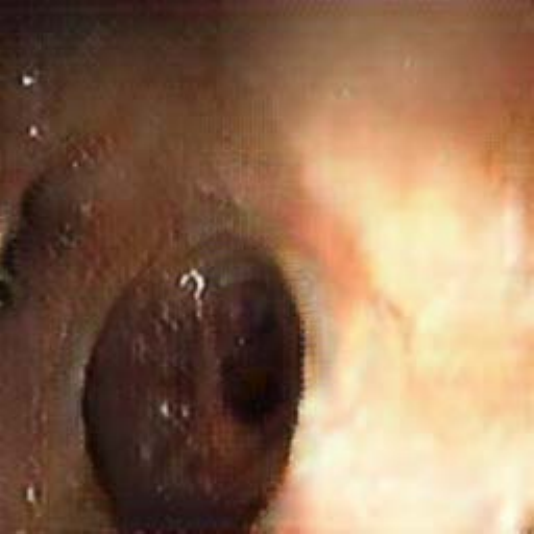}&
\includegraphics[width=0.09\textwidth]{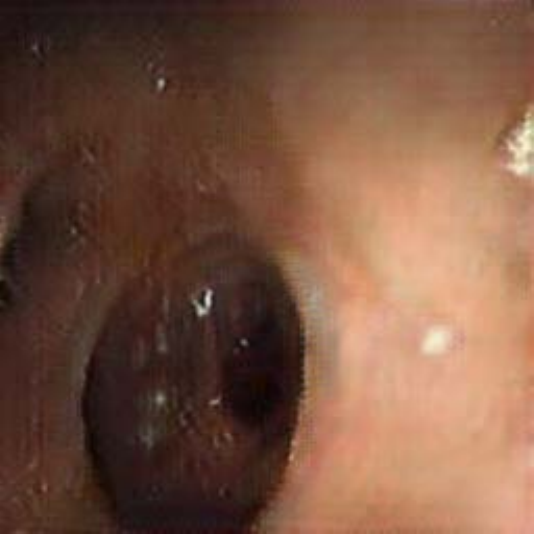}&
\includegraphics[width=0.09\textwidth]{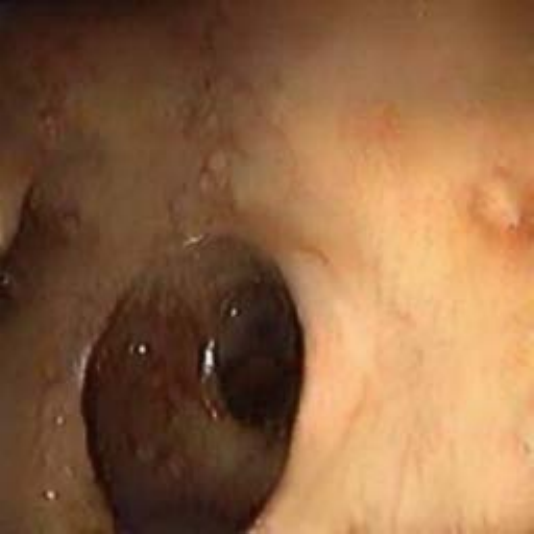}&
\includegraphics[width=0.09\textwidth]{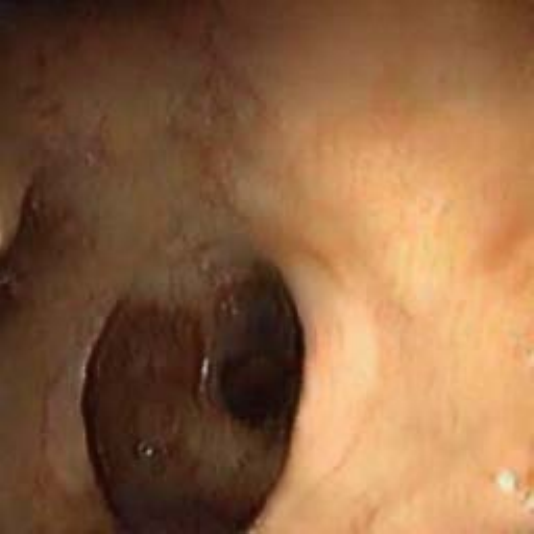}\\

\includegraphics[width=0.09\textwidth]{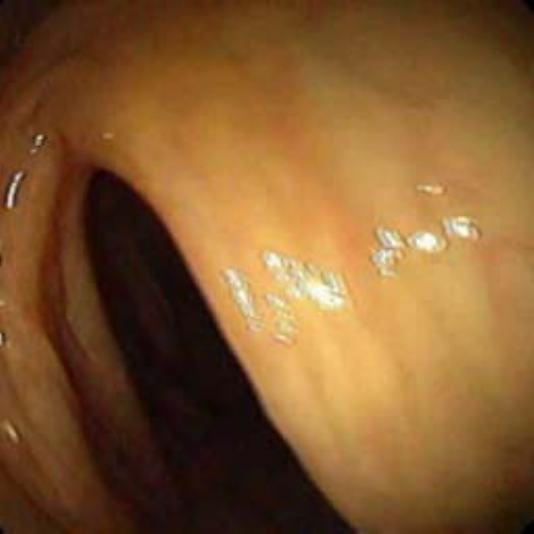}&
\includegraphics[width=0.09\textwidth]{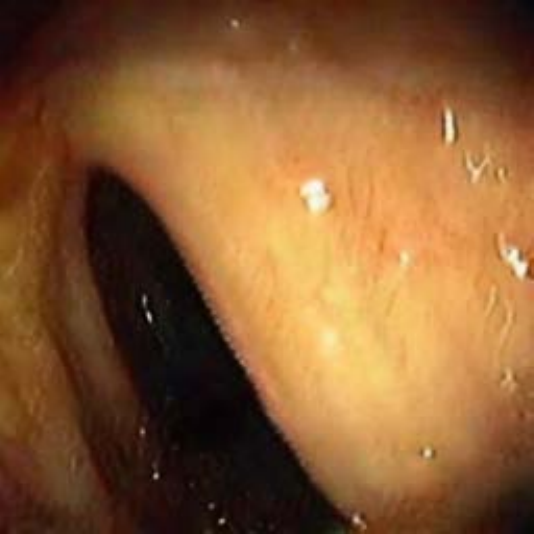}&
\includegraphics[width=0.09\textwidth]{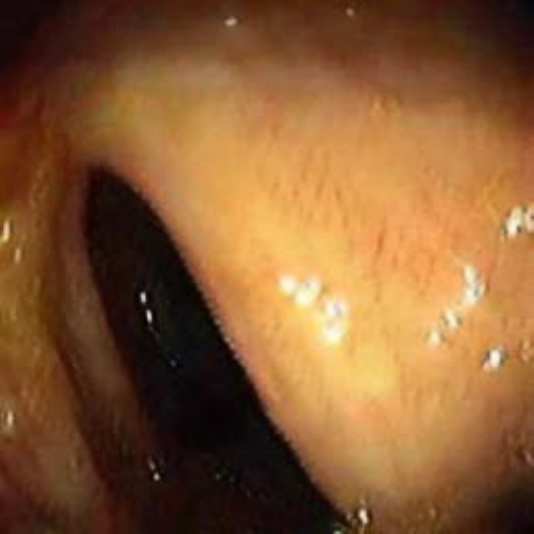}&
\includegraphics[width=0.09\textwidth]{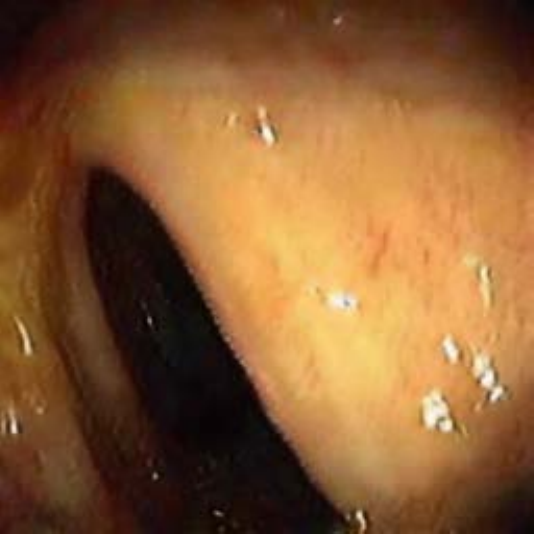}&
\includegraphics[width=0.09\textwidth]{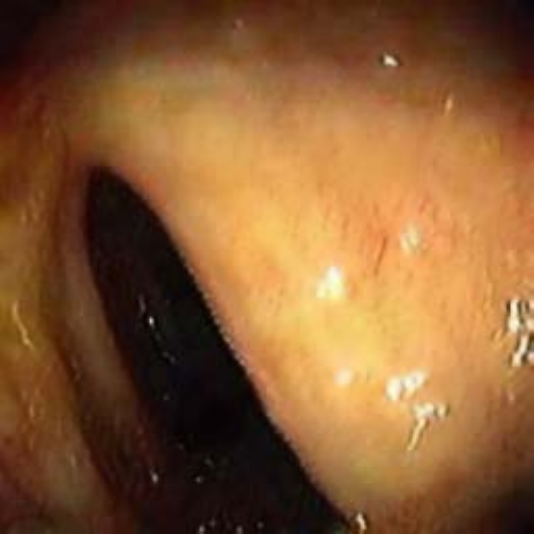}\\

\includegraphics[width=0.09\textwidth]{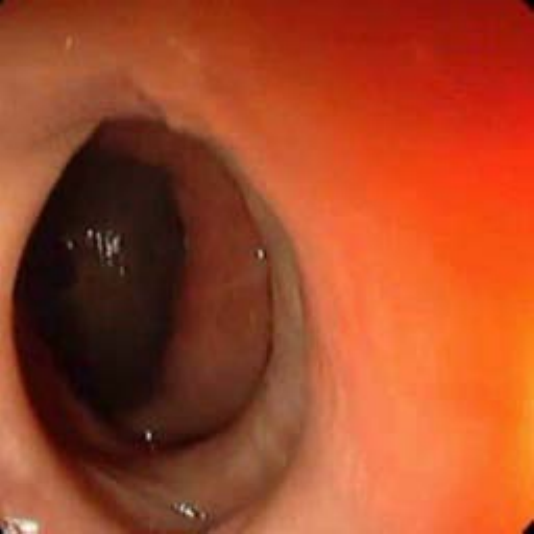}&
\includegraphics[width=0.09\textwidth]{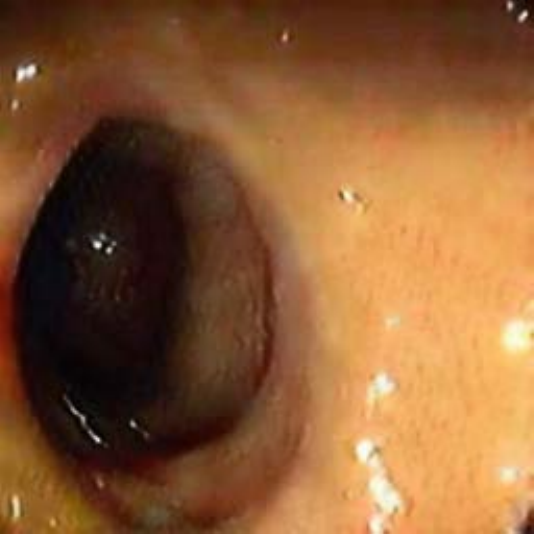}&
\includegraphics[width=0.09\textwidth]{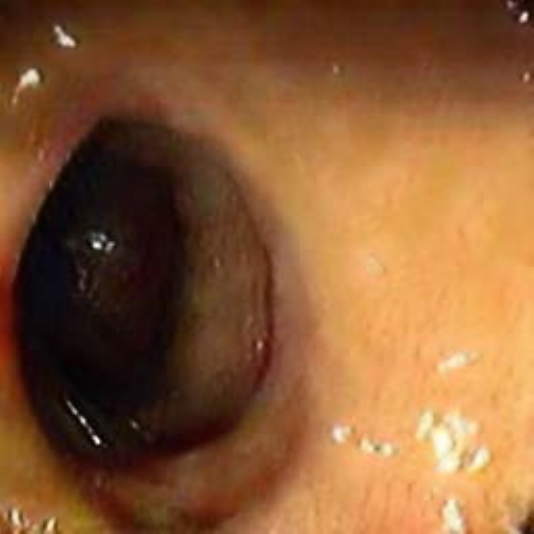}&
\includegraphics[width=0.09\textwidth]{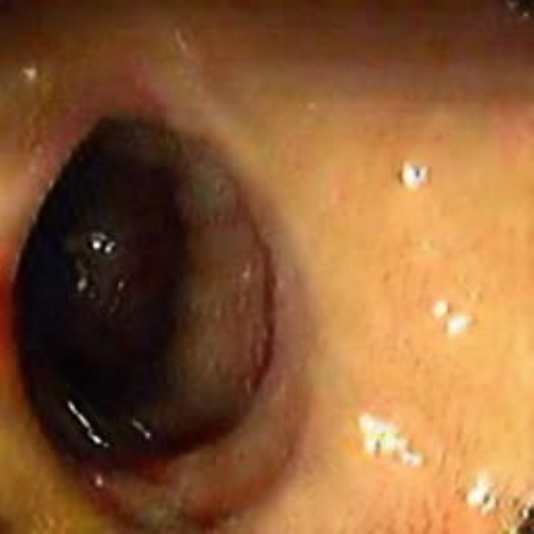}&
\includegraphics[width=0.09\textwidth]{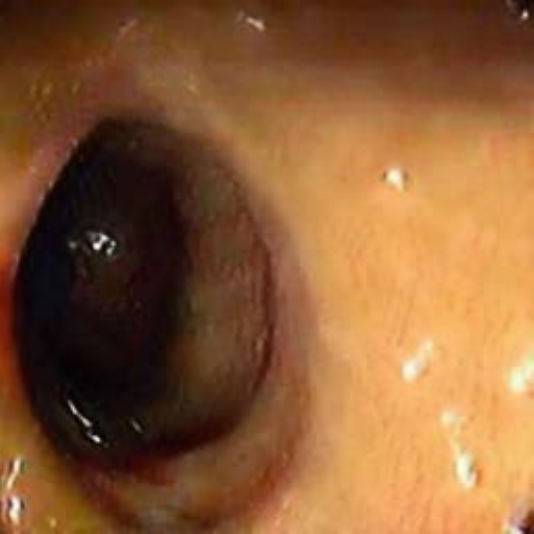}\\
\end{tabular}
\caption{The first column is input for generating OC images with different lighting, specular reflections, and textures. The first two rows show VC to OC translation. The last two rows show our model results for OC to OC.}
\vspace{-1em}
\label{fig:noise}
\end{center}
\end{figure}

We added a noise loss to our model to generate realistic OC images from given OC or VC images. The generated images have the same underlying geometry as the input image but vary in lighting, specular reflections and texture, as shown in Figure \ref{fig:noise}. The first two rows show VC to OC images. The last two rows show results for our OC to OC mapping. Just like the OC input, the model generates different lighting, specular reflections and texture. Note that the texture changes are more subtle than the changes in specular reflection and lighting. This will be addressed in future work. 

\section{Limitations and Future Work}
The missed polyps and anomalies are mostly occluded by the haustral folds. Even though our model in general works well for deep as well as shallow haustral folds, there are instances where only a partial missed surface area is highlighted for deep folds. This is understandable given the fact that we are not taking into account additional information, such as the overall colon topology. In the future, we will incorporate this information by inferring the colon centerline to improve the overall performance.

The current OC image generation creates a rather sparse distribution of images, especially in the texture space. To improve this, we will split texture, specular reflection, and lighting into three separate noise vectors which will provide finer control over these aspects and can potentially force the model to generate a more diverse set of OC images.

\section{Compliance with Ethical Standards}
This research study used retrospective human subject (OC and prior VC) data with IRB approval. In addition, OC video sequences and mesh results for Ma et al. \cite{ma2019real} were used with their permission and had prior IRB approval.

\section{Acknowledgement}
We would like to thank Dr. Sarah K. McGill (UNC Chapel Hill) for granting access to the OC videos and to Ruibin Ma for comparisons. This research was funded in part through the NIH/NCI Cancer Center Support Grant P30 CA008748 and NSF grants CNS1650499, OAC1919752, and ICER1940302.


\end{document}